\documentclass[11pt,a4paper,oneside]{article}

\pdfoutput=1

\usepackage{amsfonts}
\usepackage{amsmath}
\usepackage{amssymb}
\usepackage{float}
\usepackage[a4paper]{geometry}
\usepackage{a4wide}
\usepackage{ifthen}
\usepackage{fancyhdr}
\usepackage{color}
\usepackage{multirow}
\usepackage{textcomp} 
\usepackage{graphicx}

\usepackage{epstopdf}
\usepackage{wasysym}
\usepackage{subcaption}
\usepackage{hyperref}

\usepackage[toc,page]{appendix}

\begin{document}

\title{Extraterrestrial artificial particle sources. Application to neutrino physics and cosmic rays studies.}
\author{Nikolai Zaitsev\footnote{INNOVAEST.ORG. Correspondent e-mail: nikzaitsev@yahoo.co.uk}}
\date{20 Feb 2020}

\maketitle

\begin{abstract}
The memo is exploring possibilities to set up extraterrestrial experimental facilities to study particles physics. The Moon is considered as the most promising location for artificial particle sources outside the Earth. This natural satellite is surrounded with deep vacuum, is at low cryogenic temperatures and is always facing the Earth with one side. These features can be exploited by setting up lunar neutrino factory, which may create a possibility for more precise measurements of oscillations and possibly mass of neutrinos. Various types of facilities are discussed with focus on lunar linear accelerators and nuclear reactors. The other types such as lunar colliders or even orbiting sources are briefly mentioned too. Lunar particle accelerators pointing to Earth can also be used to calibrate atmospheric shower models, which are the key part of cosmic rays research. 
\end{abstract}

\newpage

\tableofcontents

\newpage

\section{Introduction}

\subsection{Moon-Earth neutrino long-base experiments}
\label{sec:neutrinoexperiment}

Main focus of this memo is discussion of particles source facility to set-up at the Moon. This source is proposed to produce neutrinos for very long-base experiment although one can find other applications as well. Lunar environment providing unique conditions for accelerator setup is described. Two types of experiments measuring neutrino oscillations and its mass are evaluated. Various particle sources, like particle accelerators or reactors, for experiments performed on the Moon or in the Space are mentioned.

The below should be considered as a high level description of ideas for potential extraterrestrial particle experiments. It is believed that when such projects become a reality more ideas will come. Nowadays many projects about extraterrestrial colonization are generated on the global scale. Government agencies and private companies are challenging these goals in competitive manner. So, it is reasonable to assume that the idea to build lunar particle accelerator or nuclear reactor for scientific purpose is not very remote.

\subsection{Past and existing space projects in particle and astro-physics}

The proposal to setup accelerators outside the Earth is not the first of its kind and must be considered within the global trend of space exploration by humans. 

BEAR (Beam Experiment Aboard a Rocket) project~\cite{BEAR} is the very first known accelerator in space. The respective team designed and actually launched Neutral Particle Beam linac into the space. It is reported~\cite{NPB} that this project has produced 1 MeV of neutral hydrogen beam at the height of 100 km above the Earth.  RFQ-cavities, the heart of accelerator, weighted only 55 kg \cite{BEARCERN}.

Another example is DALI ("The Dark Ages Lunar Interferometer") project, which was proposed in 2007 \cite{DALI}. This is "a Moon-based radio telescope concept aimed at imaging highly-redshifted neutral hydrogen signals from the first large scale structures formed during the Universe’s “Dark Ages” and “Epoch of Reionization". The telescope will comprise of array of antennas distributed over round area with 50 km diameter. This project is still in the stage of technical development and is supported by NASA.

Costs associated with similar proposals are still high. However, if set-up the Linear Accelerator on the Moon (LAM) should be the cheapest of its kind because:
\begin{itemize}
    \item The Moon is the closest object to Earth; 
    \item It has no atmosphere and is surrounded by deep vacuum and, therefore, does not require complex vacuum pumping techniques;
    \item The temperature in the shadow around poles riches cryogenic levels, making it possible to use energy-saving technologies;
    \item The Moon is always turned towards the Earth with one side\footnote{This occurs due to tidal locking effect}. This feature is exploited in Moon-to-Earth experiments.
\end{itemize}
Let us discuss these properties in their order.

\subsection{Lunar environment}
\label{sec:lunarspecs}

Modern acceleration technique requires low temperatures and deep vacuum. These are naturally present on the Moon for free.

On a surface of the Moon temperatures vary. It ranges from $-137$~\textdegree~C during the night to $+120$~\textdegree~C during the daylight. However, because of deep vacuum, simple artificial shadow system, like tranches, roofs and other screens will be effective measures to keep temperatures low always. It is reported by the Reconnaissance Orbiter (LRO) \cite{LRO} that some places around lunar poles~\footnote{lunar poles will be mentioned later in the search of optimal position for accelerators} are permanently in shadow and the temperatures there can reach -249\textdegree~C (or $+27$\textdegree~K). Such cryogenic temperatures help to transfer the electricity without dissipation.

The Moon has no atmosphere. For comparison, the vacuum at the surface is deeper than that inside the LHC vacuum tube by an order of magnitude. This allows better beam collimation and quality.

Deep vacuum and cryogenic temperatures allow for light design of experiments requiring much less energy budget compared to one at Earth due to no dissipation and no need for vacuum pumping. 

Likely, Lunar quakes will pose the largest problem for any setup on the Moon. They happen quite regularly~\cite{MOONQUAKE}. Four types of quakes are known: 1) caused by meteorite crashes, 2) deep-moon (700 km deep) caused by tidal forces between the Earth and the Moon, 3) thermal quakes caused by heating-cooling cycles coming from day-night changes (every $\approx29$ earth days), and finally, 4) shallow (20-30 km deep) quakes which are the strongest, up to 5.5 Richter scale, and are long-lasting for hours. Quakes is a major concern and 1) will require additional knowledge whether they evenly occur over the satellite and 2) as result, special measures to protect any lunar constructions have to be taken. 

Although the construction of accelerators can be light-weight the protection against secondary radiation induced by accelerated particles will be still needed.

All these measures will require experimental and simulation studies to optimize final design of the potential experiments and equipment.

\section{Lunar neutrino experiments}
\label{sec:lamtoearth}

\subsection{Moon-Earth experiment}
\label{sec:moonexperiment}

Moon is always turned to Earth by one side. Hence, our planet is always pinned to one point of the lunar sky. Therefore, the neutrino beam should be set-up at the border between the Front and the Far (Dark) sides of the Moon, where the Earth is visible at the horizon. Our planet occupies about 1.9 degrees of lunar sky. Orbital movement of Moon translates into back and forth movement of Earth with respect lunar reference frame. Therefore, movement of detector as a target should not exceed 2 degrees in angular size and can be targeted by the beam smoothly.

Moon has elliptic orbit around the Earth with distances, $L_{\leftmoon}$, between 363,104 and 405,696~km. In addition, Earth rotation adds another 6371 km. Detector traveling between these point will perform measurement at different distances. This corresponds to the maximum variation of source-to-detector distance of ${\sim}55,000$~km with relative change of about $14.4$~\% of the entire distance. North and South poles at Earth experience minimal distance variations and slowly change their position on lunar sky with 1 month period. This implies that, for example, existing IceCube detector~\cite{IceCube} in Antartica will be moving with minimal angular speed across lunar sky.

\subsubsection{Artificial sources of neutrino}
\label{sec:sources}

\paragraph{Accelerators}
\label{sec:accelerators}

Most of the existing accelerators use Radio-Frequency (RF) acceleration technique. They permit high energy and high intensities, however, with complex multistage system of intermediate accelerators, which accumulate particles and inject them by bunches into the main accelerator. These systems are large in size, use a lot of energy and as result are costly.

Recent advances in Plasma and Laser Wakefield Acceleration techniques (PLWA), \cite{LASERNEUTR} and \cite{LASERPOSITR}, would allow more simple and compact architecture. 
PLWA's are compact, sometimes table scale devices, with expected high energy, high luminosity, excellent collimation and quasi-mono-energetic beams. Due to their size, PLWA has a better chance to be transported to the Moon and to be setup there. 

PLWA uses plasma wakefield formed by initially accelerated electron, proton beams or laser photons. Such wakefield is used to boost charged particles including (beta-)ions (see  \cite{AWAKE} and \cite{JSATO}). PLWA's are developed with ambition in mind to achieve multi-TeV energies and beyond. However, neutrino experiments proposed below require the energy of few tens of GeV at maximum.

As it is already discussed, deep vacuum is required to maintain stability of the beam. Significant energy resources are spent to function vacuum pumps. This comes for free at Moon. Both, RF accelerators and PLWA, share the same features like pion beam formation \cite{ACCNEUTBEAM}.

\paragraph{Reactors}
\label{sec:reactors}

Reactors produce electron anti-neutrinos of low energy spanning from zero to few MeV at very high intensity. They are very attractive for long-base experiments discussed here. Nevertheless, there are negative points to operate it on the Moon. For example, nuclear reactor is active device, which is difficult to shut down when problems occur, and, therefore, it requires a lot of care. Cooling system uses water (or other liquid), which circulates  under high temperature and pressure. Additional problem is energy dissipation in vacuum which is also complicated. Therefore, reactor operation must be more complex than that at the Earth. The level of complexity of usage of nuclear reactors in space is difficult to evaluate in such short memo, however, such projects already exist since about 1970~\footnote{For example, soviet TOPAZ projects or even nuclear propulsion engines}. Another potential problem, Moonquakes, adds more requirements to safety. Overall conclusion is that the construction of nuclear reactors on the Moon is complex but feasible.

\subsection{Neutrino oscillations}
\label{sec:neutrino}

\subsubsection{Basic idea}

Oscillations are result of the interference of mass and flavour diagrams. Standard Model embeds three lepton generations, where their interactions are described with weak currents proportional to neutrino mixing matrix~\cite{NEUTRINOMIX}. 

Neutrino beams produced in mixture from neutron, pion (kaon) or muon decays of known intensity at one end (at Moon) are registered by detectors at the other end (at Earth). During flight superposition of mass eigenstates evolves. This can be registered through the measurement of flavour eigenstates as function of ratio between distance and energy of neutrino, $L/E_{\nu}$. Event rate of flavour registered in detectors while conditioned on the initial tagged flavour state is proportional to survival probability  ($P_\text{surv}$), which are driven by similar sinusoidal term:
\begin{equation}
\label{eq:oscillation}
\sin^2 \Big(\frac{1.267 \cdot \Delta m^2_{ij} \cdot L_{\leftmoon}}{E_\nu}\Big)
\end{equation}
where $\Delta m^2_{ij}$  mass squared difference for neutrino mass eigenstates, $E_\nu \text{ (GeV)}$ is energy of neutrino and $L_{\leftmoon} \text{(km)}$ is distance between the Moon and the Earth. Angles coming from parameterization of Pontecorvo-Maki-Nakagawa-Sakata (PMNS) mixing matrix~\cite{NEUTRINOMIX} govern the amplitude of rate of observed events.

\begin{figure}[ht]
\centering
\includegraphics[width=0.8\textwidth]{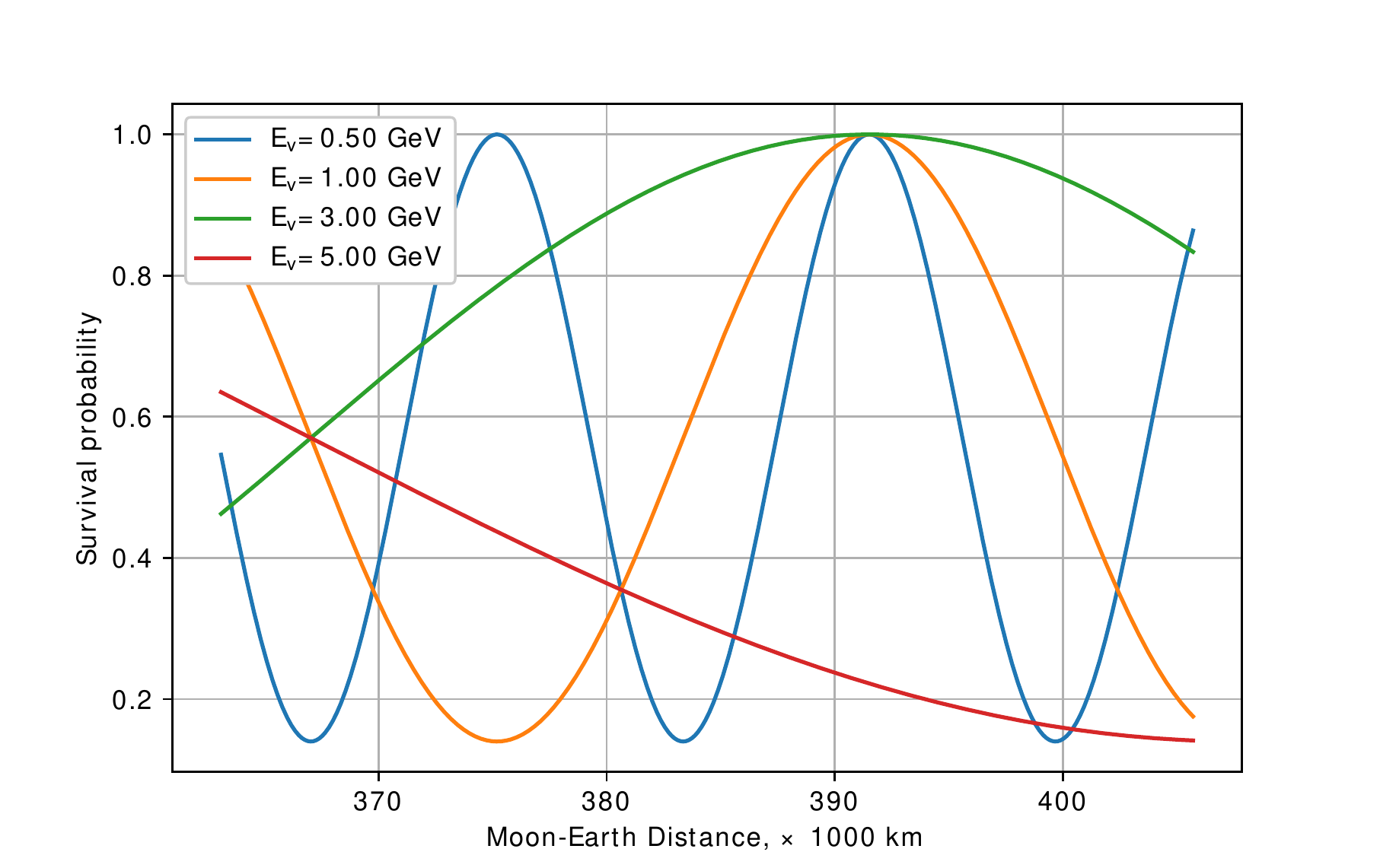}
\caption{The $\nu_{e}(\bar{\nu_{e}})$-survival probability as a function of distance, $L_{\leftmoon}$. $\Delta m^2_{12}=7.6\cdot10^{-5}~\text{eV}^2$ and $\sin^2 (2\theta_{12}) = 0.86$, see \cite{PDG}. Neutrino energies are assumed $E_{\nu} =$~0.5, 1, 3 and 5~GeV.}
\label{fig:oscillations}
\end{figure}

Fig.\ref{fig:oscillations} shows example oscillations between flavour-eigenstates of electron neutrino:
\begin{equation}
    P_\text{surv}(\nu_e \to \nu_{\bar{e}}; L_{\leftmoon}/E_\nu) = 1-\sin^2(2\theta_{12})\cdot \sin^2\Big(\frac{1.267 \cdot \Delta m^2_{12} \cdot L_{\leftmoon}}{E_\nu}\Big)
\label{eq:prob_survival}
\end{equation}
where $\sin^2(2\theta_{12})=0.86$ and $\Delta m_{12}^2 = 7.6 \cdot 10^{-5}$~eV$^2$ are chosen. The plot is made over the distance between the Moon and the Earth, $L_{\leftmoon}$, with fixed energies of neutrino: $E_\nu=0.5$, $1, 3$ and $5$~$\text{GeV}$. Oscillations due to $\Delta m_{13}^2 \approx \Delta m_{23}^2 = 2.5 \cdot 10^{-3}$~eV$^2$ require larger energy by factor of 33 to produce similar dependency in $P_\text{surv}(L_{\leftmoon})$.

\subsubsection{Precision of measurement}

\paragraph{Estimations of event rate}
\label{sex:hitrate}

Let us define event rate ($R_{ev}$) of accelerator neutrinos at detector as:
\begin{equation}
    R_{ev} = f \times \sigma \times T
\end{equation}
where $f$ is neutrino flux at detector, $\sigma$ is cross section of interaction of neutrino with detector through $\nu + N \to \mu + X$ process and $T$ is total number of protons and neutrons in the target. For on-axis neutrino beam of  $E_{\nu}=10~\text{GeV}$ it is estimated that $f \sim 10^{-15}~\nu/(\text{cm}^2\cdot\text{POT})$. Then, with $\sigma \sim 0.7 \cdot 10^{-38} \text{cm}^{-2} (E_{\nu}/\text{GeV})$ and $T \sim 6 \cdot 10^{38}$ (e.g. water detector of $1~\text{km}^3$ or $10^9~\text{kt}$) the event rate is $R_{ev} \sim 4.2\cdot 10^{-14}~\nu/\text{POT}$. Annual intensity of proton beam of $10^{20}~\text{POT}/\text{year}$ gives:
\begin{equation}
R_{ev} \sim 4.2\cdot 10^{6}~\nu/\text{year}
\end{equation}
Neutrino beam with energy $3~\text{GeV}$ will produce $4.2\cdot 10^{4}~\nu/\text{year}$ events. See details of flux calculations in Appendix~\ref{sec:appendix}.

This estimation is really rough and is aimed to illustrate only feasibility of such project. Some optimal experiment design has to be found as a result of interplay between statistics and resolution of measured physics parameters. Estimated event rate does not include detection probability. Let's use this information further.

\paragraph{Distance and Energy resolution}

Measurement of oscillations requires knowledge of distance, $L_{\leftmoon}$, and  energy $E_{\nu}$.

Any moment in time, the distance between centers of the Moon and the Earth, $L_{\leftmoon}$, can be known with precision up to 1 mm (or with relative error $2.5\cdot 10^{-12}$) and hence is assumed to be precise. Such distance measurement is possible due to the existence of several retro-reflectors placed on the Moon, which are used to measure and monitor the movement of the satellite.

\begin{figure}[ht]
\centering
\includegraphics[width=1.0\textwidth]{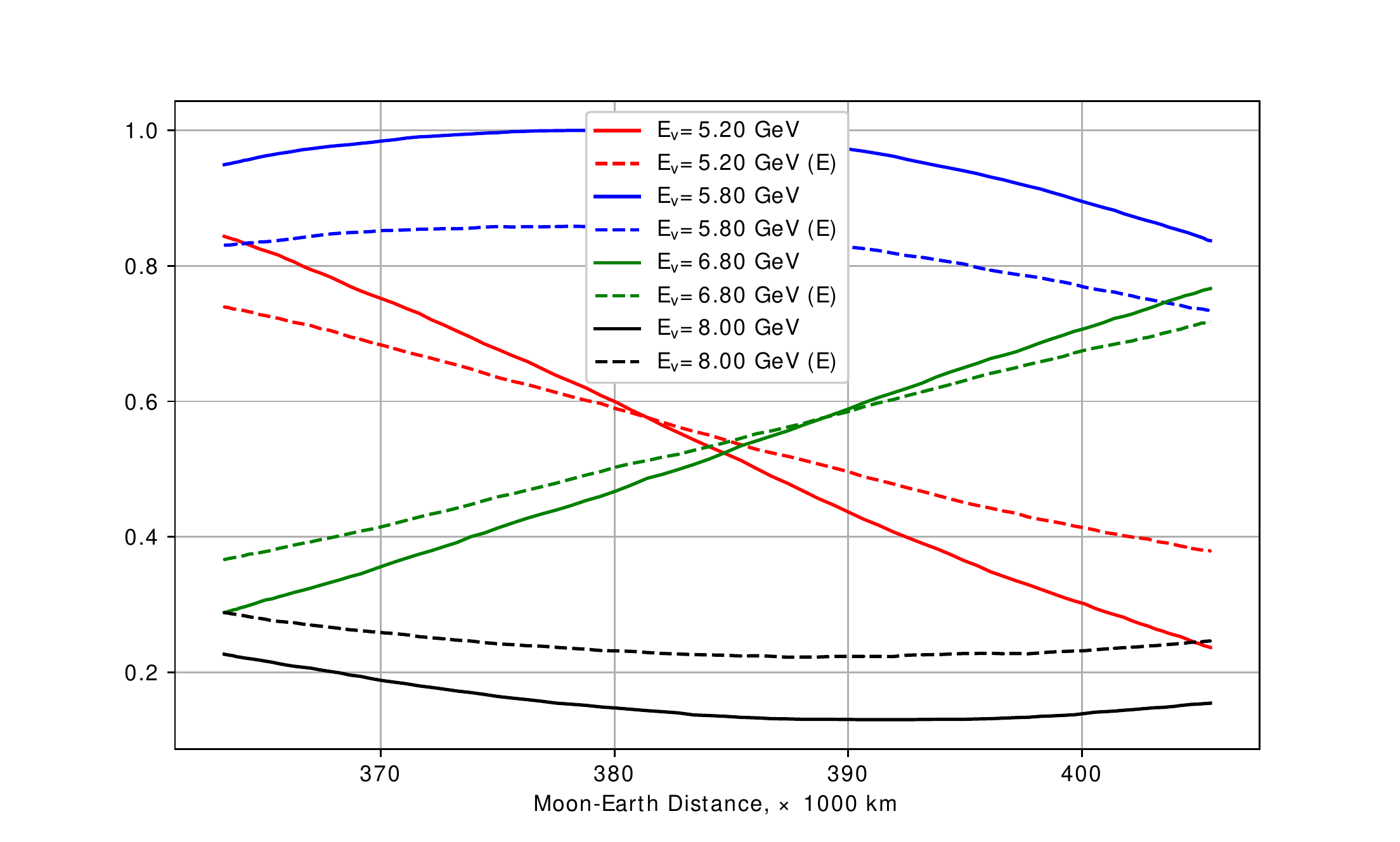}
\caption{The $\nu_e$($\bar{\nu_e}$) survival probability as a function of distance, $L_{\leftmoon}$. Theoretical and 'measured' (marked with '(E)') probabilities are presented. The same physics parameters are taken as were used in Fig.~\ref{fig:oscillations}. Neutrino energies shown are $E_{\nu} = 5.2,~5.8,~6.8,~8.0$~GeV. Similar choice should maximise the measurement precision.}
\label{fig:oscillations500mev}
\end{figure}

For the illustration of effect of energy resolution, toy Monte Carlo is built to simulate 10k neutrinos with energy normally distributed and centered at $E_\nu=\{3,~5,~10\}$~GeV with precise energy measurement and then with measurement uncertainty of $\Delta E_{\nu} \approx \{0.2, 0.35, 0.7\}$~GeV~\footnote{Calculated with $\Delta E_{\nu} / E_{\nu} = 0.070 + 0.033/\sqrt{E_{\nu}}$ (from \cite{DUNEDE}), where $[E_{\nu}]=[\text{GeV}]$}.
Fig.\ref{fig:oscillations500mev} shows how original oscillation pattern dilutes due to energy measurement error, where the amplitude of oscillation is sensitive to $\sin^2(2\theta_{12})$ and the period of oscillation is sensitive to $\Delta m^2_{21}$. 

Narrow band energy neutrino beam will improve this result:
\begin{itemize}
    \item Off-beam neutrinos have narrow energy bandwidth, however, at the cost of production intensity.
    \item Magnet system can select specific energy band. Intensity in this case is even smaller than in the previous one;
\end{itemize}

Of course, narrow band energy neutrinos are registered with the same energy resolution as mentioned above, but in such experiments the signal events can be selected with energy around the known one, thus reducing the background neutrinos. If done, the dependency of survival probability as function of distance to the Moon will show more pronounced oscillations and, therefore, will be more sensitive to $\Delta m$'s.


\subsubsection{Experimental aspects and systemic effects}

The existing long-baseline experiments share similar systemic problems:
\begin{itemize}
    \item the necessity to normalize "near" and "far" fluxes of neutrinos;
    \item the necessity to normalize fluxes and cross section at different energies. This is needed to correctly translate visible event rate into survival probability and then into the measurement of oscillations.
\end{itemize}
Same problems exist for experiments with accelerator and nuclear sources. Moon-to-Earth experiment will be free from such systematic errors because oscillation scan can be done with neutrino beam with same energy spectrum, while the target (Earth detector) moves alongside the beam. Moreover, in this case the measurement of $\Delta m^2$ does not require absolute normalization, while for the measurement of PMNS-matrix elements it is still required.

Through the formula (\ref{eq:prob_survival}) one can see that the precision of $|\Delta m^2|$ measurement is sensitive to the precision of $E_\nu / L_{\leftmoon}$ measurement. Keeping that in mind one can use Table 14.4 in \cite{PDG} as a reference. This table quotes $|\Delta m^2|$-sensitivity potential as ratio  $E_{\nu}/L_{\leftmoon}~\text{(MeV/m)}$, which indicates that the lunar experiment has sensitivity to $|\Delta m^2|$ at $\sim10^{-5}$~$eV^2$. This level of precision is next to solar neutrino studies, which have best potential sensitivity. However, although Sun produces huge amount of neutrinos, they have low energy and more difficult to detect, their flavour mix is not known at the source and have broad energy spectrum. Respective results have systematic error due to (solar) model dependency. The same is true about atmospheric neutrinos as one gets dependent on cosmology physics models. This limits the sensitivity to sinusoidal structure of oscillations measured from solar or atmospheric sources.

\subsection{Neutrino mass measurement with time of flight technique}

\subsubsection{Basic idea}

Another possible application of neutrino beam directed to earthly detector is the measurement of neutrino mass by using Time of Flight (ToF) technique. This technique applied to neutrino was first proposed by G.Zatsepin in 1968 \cite{ZATSEPIN} to measure time-energy dependence of these particles emitted by supernovae. This idea led to observation of SN1987a~\cite{SN1987A} neutrinos and the measurement of upper limit on neutrino mass.

\begin{figure}[ht]
\centering
\includegraphics[width=0.8\textwidth]{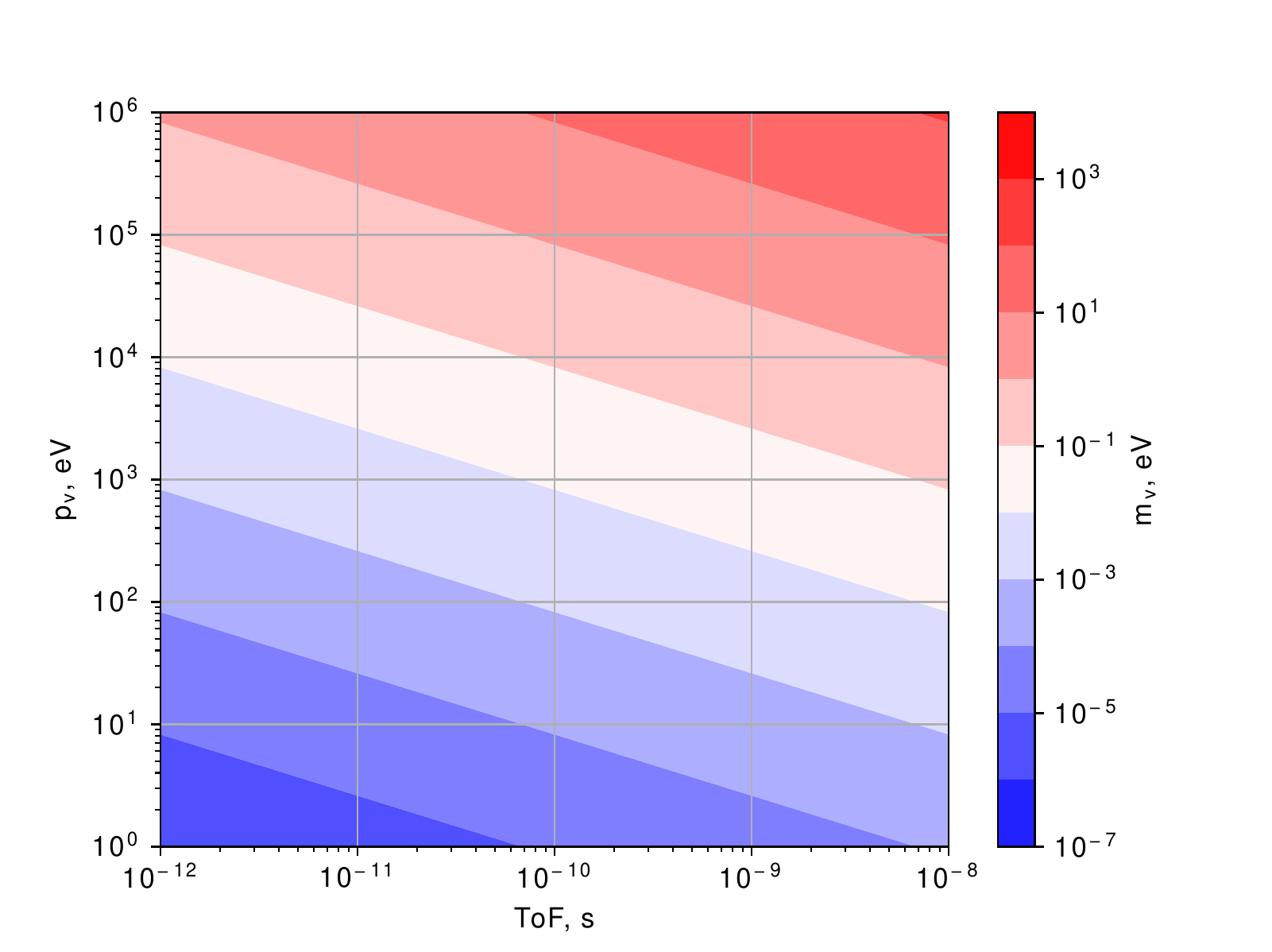}
\caption{Scatter plot shows relationship between ToF (X-axis), momentum (Y-axis) and mass of neutrino (color map).}
\label{fig:tof_mass_scatter}
\end{figure}

The same idea can be used to measure ToF difference between neutrino and synchronized at source photon, $\delta t_{\gamma,\nu} = t_{\gamma} - t_{\nu}$. This allows the measurement of the mass of neutrino as:
\begin{equation}
\label{eq:neutmasstof}
    m_{\nu} = \frac{p_{\nu}}{c}\cdot\sqrt{\Big(\frac{\delta t_{\gamma,\nu} \cdot c}{L_{\leftmoon}}+1\Big)^2-1}
\end{equation}
By changing momentum, $p_\nu$, and $t_{\gamma,\nu}$ one can estimate the limits of mass measurement at Moon-Earth distance. Such scatter plot is shown in Fig.\ref{fig:tof_mass_scatter}, where neutrino with $p_{\nu}=1$~keV, $L_{\leftmoon}(max)=405,696$~km and delay of $\delta t_{\gamma,\nu} \sim 1$~ns will produce the limit on neutrino mass estimation at $m_\nu \leq 38.7$ meV. The same limit can be achieved for 10 keV neutrino, but in this case the time precision must be at $\delta t_{\gamma,\nu} \sim 10$~ps. Therefore, ToF measurement of neutrino mass puts very strict constrains on detector properties, which has to be sensitive to neutrinos with $\sim1~\text{keV}$ momentum and with time resolution of the order of nanoseconds to achieve mass limit of $0.1~\text{eV}$. Plastic scintillators are capable to respond to the electron within few hundreds picoseconds, however, they are not that sensitive to the keV electrons. Some hope can be attributed to germanium detectors which demonstrate sensitivity to neutrinos in sub-keV region and give time resolution at microseconds scale~\cite{GERMANIUM}.

Source of low energy neutrinos is also a challenge because these neutrinos should be time tagged at the source. It can be done by measurement of time difference between leading photon and delayed neutrino, $\delta t_{\gamma,\nu}$, which are produced within same bunch. One example is Laser wakefield accelerators emitting beta-ions and laser photons in synchronicity. Another example is nuclear reactor equipped with selection of monochromatic neutron beam and Fermi-chopper~\cite{TOFREACTOR} to strobe neutrons.

High time precision involving leading photon might require some interferometer schema (see \cite{FEMTOSEC}). It can have certain complications for time periods when detector is on the other side of the Earth and is not directly visible from the Moon. Small satellites with reflectors positioned at Lagrange points within Earth-Moon system could solve this particular problem. 

Nuclear reactors produced neutrinos at high rates of $10^{20}~\bar{\nu_e}/\text{s}$ per $1~\text{GWt}$ of power. With acceptance of $1/(4\pi\cdot L_{\leftmoon})^2\sim 4\cdot10^{-20}$ one can expect only 4 neutrinos to pass through $1~\text{m}^3$ detector per second or $\sim10^{8}~\bar{\nu_e}/\text{year}$. Only fraction of these from $\sim0.1-5~\text{keV}$ energy band can be used in this ToF experiment. Estimation of respective flux is obscured by the absence of measurements of reactor flux in area of sub-MeV neutrinos. Additionally, at these energies cross section of neutrino interaction with detector material is low. Nevertheless, the feeling is the flux arriving at Earth should be very small. 

These arguments suggest the use of accelerator technique as a better alternative. For example, plasma/Laser wakefield accelerators could serve this purpose. But in general, it is likely that such sources do not exist, because of natural limitations of neutrino production as they have small mass (from beta-decay) or because slow protons do not have enough energy to generate pions. The expectation is that this proposal for neutrino mass measurement with ToF technique will produce respective ideas about how to produce highly intense beam of slow neutrinos. 

\subsection{Other experiments}

\subsubsection{Calibration of cosmic rays physics models}

Lunar accelerator can produce particles of known ID (electrons, positrons, protons, neutron, neutrinos or ions) and with energy of known profile or even mono-energetic one. By shooting them into Earth one can better study the physics of atmospheric showers and calibrate respective models \cite{COSMICRAYS} which are key part of cosmic rays physics.

For example, Laser induced electron beam can have an energy of 0.01-1 GeV. $3$~mrad beam will create 1200 km radius spot on Earth surface. Of course, beam intensity of $10^{15}-10^{20}$ electrons per second will be extremely high for the living species, but it can be lowered down to the level sufficient to calibrate the models.

\subsubsection{Lightnings as spark chamber} Another interesting potential application of particle accelerator with electrons is the existence and progressing development of "World Wide Lightning Location Network" \cite{WWLLN}. Again, by shooting charged particles at locations with potential lightning can give us a possibility to model and calibrate the response. If done so, this network can help to develop better lightning models used as a global spark chamber with capacity still to explore for high energy and cosmic rays physics. 

\subsubsection{Accelerator and detector on the Moon}

Any other setup will require detectors placed locally at the Moon. Note, that since the Moon is in the vacuum there is no need to make detector setup compact which is required at Earth due to space limitations in the experimental cave. Detectors can be positioned quite sparsely with a long arm even on purpose, for example, in order to improve spacial measurements.

Large circular collider can be placed at the bottom of the large enough crater circus. Similar artificial trench and crater boundaries will be the protection for outside lunar environment possibly inhabited with people.

\subsubsection{Orbiting particle sources}

One such experiment was mentioned at the beginning of the memo. However, it is difficult to use such device as it has to constantly point to the detector on Earth. Alternatively, accelerator can be positioned either in geo-stationary points in space or orbiting around the Moon. In the latter case the study of the Moon structure is possible through the scan with neutrinos.

\section{Conclusions}

Starting this memo author had an impression that lunar project is very remote in time and is not possible to execute. However, by making these simple estimations and observing determination of various governments and entrepreneurs this project seems to become feasible in the foreseeable future. Among many other possibilities, the Moon is quite suitable place for studies of neutrino and atmospheric showers as a link to cosmic rays physics. 

\section{Acknowledgments}

Author is grateful to Dmitry Naumov, (Dubna, JINR) for challenging results and more clear explanations about the physics. Peter Hristov is aknowledged for critical views and suggestions to look plasma/laser wakefield accelerator technologies.
My daughter Valentina helped me to streamline my English in the very first version. At the end I would like to mention that, the work on this piece has forced me to refresh some 20 years old knowledge about high energy physics which is a reward by itself. This work is written at the own initiative of author without any support.

\newpage

\begin{appendices}
\section{Flux estimation}
\label{sec:appendix}

\begin{figure}[ht]
\centering
\includegraphics[width=0.8\textwidth]{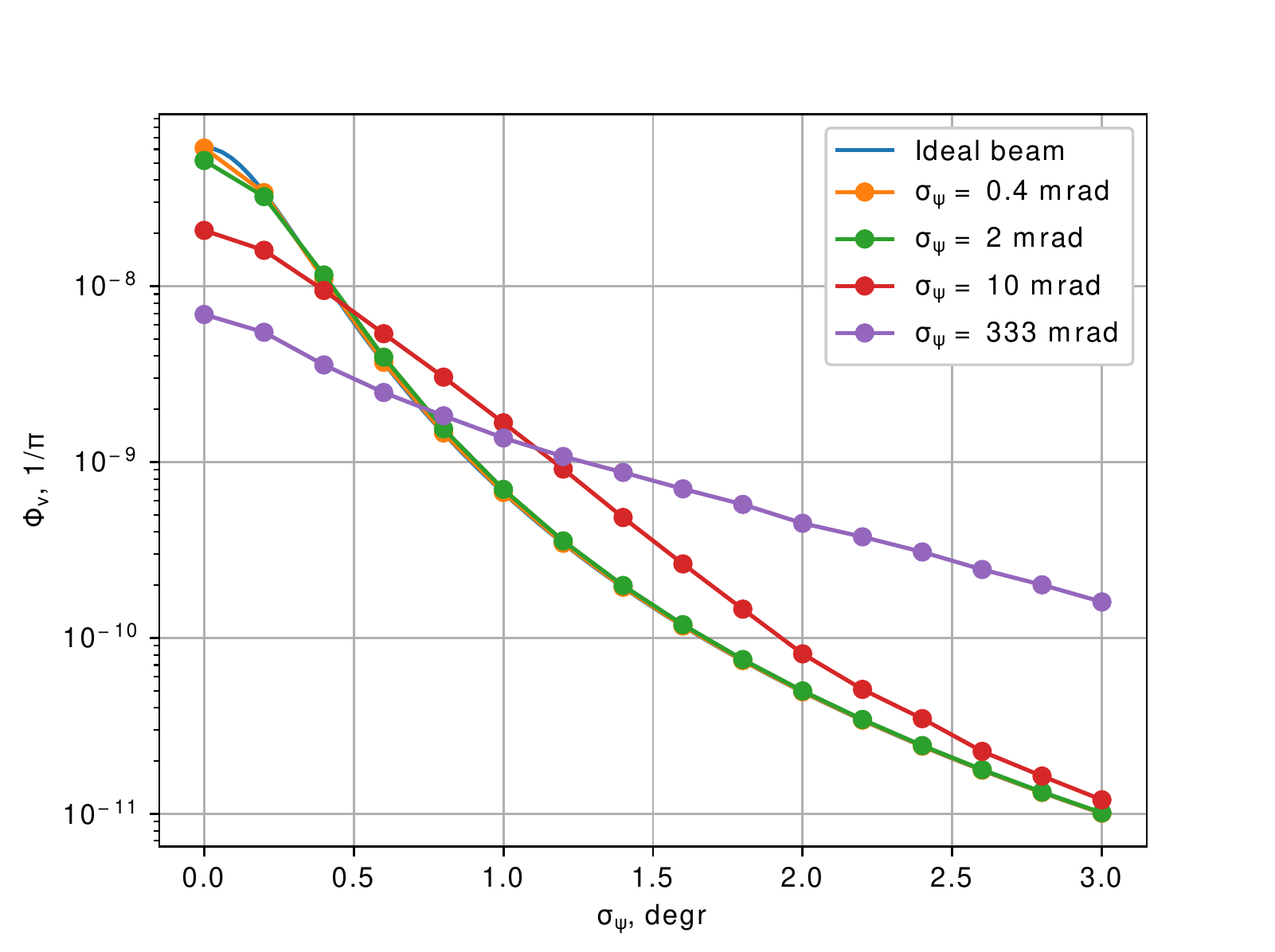}
\caption{Neutrino flux, $\Phi_{\nu}$, as a function of $\theta$ off-axis for lunar experiment. Different pion beam diversions are assumed, $\sigma_{psi} = 0,~0.4,~2,~10,~333$~mrad. Dots are connected with linear interpolation. $E_{\nu}=10~\text{GeV}$.}
\label{fig:pion_beam}
\end{figure}

Total neutrino flux at detector, $f$, can be factorised into two parts: flux of neutrino produced from single pion, $\Phi_{\nu}$, on top of flux of pions produced from single proton, $\eta(E_{\pi},\theta_{\pi})$:
\begin{equation}
f = \eta \cdot \Phi_{\nu}
\label{eq:totalflux}
\end{equation}
where $\eta$ is taken constant as an approximation. Flux $\Phi_{\nu}$ captures dependence on energy of pion decayed into neutrino and muon, size of one dimension of detector, $D$, distance to it, $L$, and off-beam axis, $\theta$:  
\begin{equation}
\Phi_{\nu} (D, L, \gamma_{\pi},\theta) = \frac{\gamma_{\pi}^2 \cdot D}{\pi \cdot L} \cdot
\int_{\theta}^{\theta+D/L} \frac{(1+\tan^2 \alpha)^{3/2}}{(1+\gamma_{\pi}^2\cdot \tan^2 \alpha)^2} d \cos \theta
\label{eq:flux}
\end{equation}
which holds for relativistic  pions with $1 \ll \gamma_{\pi}$. This relationship comes from kinematics (see (9) in \cite{ACCNEUTBEAM} or (10) in \cite{RAMM}). The kinematic relationship between energy of pion and neutrino is also used:
\begin{equation} \label{eq:energy_pion}
E_{\nu} = 
\frac{m_{\pi}^2 - m_{\mu}^2}{m_{\pi}^2} \cdot \frac{E_{\pi}}{1+(\gamma_{\pi}\theta)^2} = \frac{m_{\pi}^2 - m_{\mu}^2}{m_{\pi}} \cdot \frac{\gamma_{\pi}}{1+(\gamma_{\pi}\theta)^2}
\end{equation}
These expressions are used to extrapolate reported experimental data to estimate flux in lunar experiment. For this purpose we choose MiniBooNE~\cite{MBNE} and T2K~\cite{T2K} experiments. Their respective parameters are listed in Table~\ref{table:experiments}. 
\begin{table}[ht]
\begin{tabular}{ |c|c|c|c|c|c| } 
 \hline
 Experiment & $L$, km & $A$, $\text{m}^2$ & $\theta, \text{degr}$ & $\overline{E_{\nu}}, \text{GeV} $ & $\overline{E_{\pi}}, \text{GeV} $ \\ 
 \hline 
 MiniBooNE  & 0.56 & $\sim 100$ & 0 & 0.8 & 2.2 \\ 
 T2K        & 295  & 1600 &  2.5 & 0.6 & 1.9-5.4 \\ 
 MOON10     & $3.82\cdot 10^{5}$ & $10^{6}$ & 0 & 10 & 23.4 \\ 
 MOON3     & $3.82\cdot 10^{5}$ & $10^{6}$ & 0 & 3 & 7 \\ 
 \hline
 \hline
\end{tabular}
\begin{tabular}{ |c|c|c|c| } 
Experiment & $\Phi_{\nu}, \nu/\pi$ & $f_{\text{exp}}$,~$\frac{\nu}{\text{cm}^2\cdot \text{POT}}$ & $\eta$,~$\frac{\pi}{\text{cm}^2\cdot \text{POT}}$ \\
\hline
MiniBooNE  & $3.17\cdot 10^{-2}$& $5.19\cdot 10^{-10}$& $2.0\cdot 10^{-8}$ \\
T2K        & $5.82\cdot 10^{-7}$& $1.59\cdot 10^{-14}$& $2.7\cdot 10^{-8}$ \\
MOON10      & $6.1\cdot 10^{-8}$& - & - \\
MOON3      & $5.5\cdot 10^{-9}$& - & - \\
\hline
\end{tabular}
\caption{Summary of parameters used in calculations, where $L_{\text{det}}$ is distance to detector, $A$ is effective area of detector frontal to the beam, $\theta$ is off-axis angle, $\overline{E_{\nu}}$ is average energy of neutrino, $\overline{E_{\pi}}$ is average energy of parent pion implied with use of (\ref{eq:energy_pion}), $\Phi_{\nu}$ is flux estimated from (\ref{eq:flux}), $f_{\text{exp}}$ is flux reported by experiments and $\eta$ is pion flux estimated from (\ref{eq:totalflux}). MOON10 and MOON3 are flux projections with $E_{\nu}=10~\text{GeV}$ and $3~\text{GeV}$, respectively.}
\label{table:experiments}
\end{table} \\
MiniBooNE value, $\eta_{MBNE} = 2\cdot 10^{-8}~\pi/(\text{cm}^2\cdot \text{POT})$, is chosen further. These estimations are rough and suffer from the following assumptions related to specifics of formation of neutrino beam:
\begin{itemize}
    \item energy spectrum of pions from $p+X_{\text{fix}} \to N\pi + X_{\text{out}}$ and their angular spread are neglected and replaced with single average energy and setting $\theta_{\pi}=0$;
    \item dependency of proton-to-pion conversion on target material and energy is also neglected.
\end{itemize}
Impact of angular diversion of pion beam, $\psi$, on flux estimation is calculated. Its typical spread is about $3\sigma_{\psi} \approx 1.2$~mrad~\cite{BEAM_ANGLE}. This effect on flux is checked via simulation of pion beam with zero and non-zero diversions as a function of $\theta$ off-axis angle. For typical beam imperfections of $\sigma_{\psi}<2~\text{mrad}$ no significant impact is found. It starts to be visible at $\sigma_{\psi}>10~\text{mrad}$, see \ref{fig:pion_beam}
 \\
Therefore, total neutrino flux estimations for detector of Moon-Earth experiment are:
\begin{equation}
    f(E_{\nu} = 3~\text{GeV}) = 5.5 \cdot 10^{-9} \cdot 2 \cdot 10^{-8} \approx 10^{-16}, \frac{\nu}{\text{cm}^2\cdot \text{POT}}
\end{equation}
\begin{equation}
    f(E_{\nu} = 10~\text{GeV}) = 6.1 \cdot 10^{-8} \cdot 2 \cdot 10^{-8} \approx 10^{-15}, \frac{\nu}{\text{cm}^2\cdot \text{POT}}
\end{equation} \\
Note, that with increase of pion energy, $E_{\pi}$, the beam collimation improves with Lorentz boost quadratically.
\end{appendices}


\newpage

\begin{thebibliography}{99}
\bibitem{BEAR} "BEAR (Beam Experiments Aboard a Rocket) project". Final report. Volume 1, Project Summary. LA-11737-MS,Vol.I, BEAR-DT-7-1
\bibitem{NPB} "A Brief History of High Power RF Proton Linear Accelerators", by John C. Browne, preprint of Los Alamos National Laboratory.
\bibitem{BEARCERN} "A Linear accelerator in space· The beam experiment aboard a rocket". P. G. O'Shea, T. A. Butler, M. T. Lynch, K. F. McKenna, M. B. Pongratz, T. J. Zaugg., Proceedings of the "Linear Accelerator Conference 1990", Albuquerque, New Mexico, USA
\bibitem{DALI} "The Dark Ages Lunar Interferometer (DALI)." by Joseph W. Lazio, T \& Kasper, JÃ¼rgen \& Jones, D \& Burns, J \& Furlanetto, S \& Weiler, K \& Macdowall, R \& Demaio, L \& Bale, Stuart \& Ellingson, S \& Greenhill, L \& B. Taylor, G. (2007).
\bibitem{LRO} Published as "Nature" news. doi:10.1038/news.2009.1149
\bibitem{MOONQUAKE} "The Lunar Seismic Network: Mission Update." by Neal, Clive \& Banerdt, William \& Chenet, Hugues \& Gagnepain-Beyneix, J \& Hood, L \& Jolliff, B \& Khan, A \& J. Lawrence, D \& LognonnÃ©, Philippe \& Mackwell, S. (2004).
\bibitem{IceCube} The IceCube Collaboration. "The Design and Performance of IceCube DeepCore". arXiv:1109.6096v1 [astro-ph.IM]. 25 Sep 2011.
\bibitem{LASERNEUTR} "Neutrino oscillation studies with Laser-Driven
Beam dump facilities". by S.V. Bulanov1,5, T. Esirkepov1,6, P. Migliozzi2, F. Pegoraro3, T. Tajima1, F. Terranova.. arXiv:hep-ph/0404190v2 2 Dec 2004.
\bibitem{LASERPOSITR} "Table-Top Laser-Based Source of Femtosecond, Collimated, Ultrarelativistic Positron Beams". by Sarri, G. and Schumaker, W. and Di Piazza, A. and Vargas, M. and Dromey, B. and Dieckmann, M. E. and Chvykov, V. and Maksimchuk, A. and Yanovsky, V. and He, Z. H. and Hou, B. X. and Nees, J. A. and Thomas, A. G. R. and Keitel, C. H. and Zepf, M. and Krushelnick, K.. Phys. Rev. Lett. 110, 25, Jun2013.
\bibitem{AWAKE} "Acceleration of electrons in the plasma wakefield of a proton bunch", by E.Adli et al.. Nature, volume 561, pages363–367 (2018).
\bibitem{JSATO} "Monoenergetic Neutrino Beam for Long Baseline Experiments", by J. Sato. Phys. Rev. Lett. 95, 131804, 2005.
\bibitem{ACCNEUTBEAM} "Accelerator Neutrino Beams", by Sacha E. Kopp.. arXiv:physics/0609129v1 [physics.acc-ph], 14 Sep 2006.
\bibitem{NEUTRINOMIX} 1) B. Pontecorvo, Zh. Eksp. Teor. Fiz. 33, 549 (1957) and 34, 247 (1958). \\
2) 5. Z. Maki, M. Nakagawa, and S. Sakata, Prog. Theor. Phys. 28, 870 (1962).
\bibitem{PDG} M. Tanabashi et al. (Particle Data Group), Phys. Rev. D 98, 030001 (2018).
\bibitem{DUNEDE} "Neutrino oscillations at DUNE with improved energy reconstruction", by De Romeri, Valentina et al. JHEP 1609 (2016) 030. arXiv:1607.00293 [hep-ph] PCCF-RI-15-04
\bibitem{ZATSEPIN} "Toward the Possible Determination of an Upper Bound on the Neutrino Mass by Time of Flight", G. T. Zatsepin, JETP Letters,8 ,p. 333, 1968.
\bibitem{SN1987A} a) "Observation of a Neutrino Burst from the Supernova SIV1987A", by K.Hirata et.al. (Kamiokande II collaboration), Phys.Rev.Lett. vol.58, num 14. April 1987. \\
b) "Observation of a Neutrino Burst in Coincidence with Supernova 1987A in the Large Magellanic Cloud", by R.M.Bionta et.al (IMB collaboration), Phys.Rev.Lett. vol.58, num 14. April 1987. \\
c) "Possible detection of a neutrino signal on 23 February 1987 at the Baksan underground scintillation telescope of the Institute of Nuclear Research", Pis'ma Zh. Eksp. Teor. Fiz 45, No 10, May 1987.
\bibitem{GERMANIUM} "Germanium detectors with sub-keV sensitivities for neutrino and dark matter physics", by  Arun Kumar Soma et al 2015 J. Phys.: Conf. Ser. 606 012011.
\bibitem{TOFREACTOR} "Neutron Time-of-Flight Spectroscopy", John R. D. Copley and Terrence J. Udovic. J. Res. Natl. Inst. Stand. Technol. 1993 Jan-Feb; 98(1): 71–87
\bibitem{FEMTOSEC} "Measurement of femtosecond electron beam based on frequency and time domain schemes". K. Kan, M. Gohdo, T. Kondoh, I. Nozawa, J. Yang , Y. Yoshida. Proceedings of IBIC2016, Barcelona, Spain.
\bibitem{COSMICRAYS} PDG, Version Revised in Oct2017 by J.J. Beatty (Ohio State Univ.), J. Matthews (Louisiana State Univ.), and S.P. Wakely (Univ. of Chicago).
of Barcelona) and M. Yokoyama (Tokyo U.; Kavli IPMU (WPI), U. Tokyo)
\bibitem{WWLLN} World Wide Lightning Location Network, \url(http://wwlln.net/)
\bibitem{RAMM} "Neutrino spectra from the two-body decay of relativistic parents", by C.A. Ramm. at 4 April 1963
\bibitem{MBNE} "The Neutrino Flux prediction at MiniBooNE", by A. A. Aguilar-Arevalo et. al. (MiniBooNE Collaboration). arXiv:0806.1449v2 [hep-ex] at 15 May 2009. 
\bibitem{T2K} "The T2K Neutrino Flux Prediction", K.Abe et. al. (T2K collaboration). arXiv:1211.0469v3 [hep-ex] at 22 Jan 2013.
\bibitem{BEAM_ANGLE} "Measurement of the muon neutrino charged-current single pion+ production on hydrocarbon using the T2K off-axis near detector ND280", by K.Abe et.al. (T2K collaboration). arXiv:1909.03936v3 [hep-ex] at 17 Sep 2019
\end{thebibliography}
\end{document}